\documentclass[aip,apl,amsmath,amssymb,onecolumn]{revtex4-2}
\usepackage{graphicx}
\usepackage{dcolumn}
\usepackage{bm}
\usepackage[utf8]{inputenc}
\usepackage[T1]{fontenc}
\usepackage{mathptmx}
\usepackage{changepage}
\usepackage{booktabs}
\usepackage{textcomp} 
\usepackage{gensymb}
\usepackage{array} 
\usepackage[margin=20mm]{geometry}
\usepackage{amssymb,amsmath,amsthm,enumitem}
\usepackage[mathlines]{lineno}
\usepackage{graphics} 
\usepackage{longtable} 
\usepackage{url} 
\usepackage{bm} 
\usepackage{upgreek}
\usepackage{amsthm}

\begin{document}


\title{Numerical analysis and optimization of a hybrid layer structure for triplet--triplet fusion mechanism in organic light-emitting diodes}


\author{Jun-Yu Huang}
\affiliation{Graduate Institute of Photonics and Optoelectronics and Department of Electrical Engineering, National Taiwan University, Taipei 10617, Taiwan}
\affiliation{Cavendish Laboratory, University of Cambridge, Cambridge CB3 0HE, UK}
\author{Hsiao-Chun Hung}
\affiliation{Graduate Institute of Photonics and Optoelectronics and Department of Electrical Engineering, National Taiwan University, Taipei 10617, Taiwan}
\author{Kung-Chi Hsu}
\affiliation{Graduate Institute of Photonics and Optoelectronics and Department of Electrical Engineering, National Taiwan University, Taipei 10617, Taiwan}
\author{Chia-Hsun Chen}
\affiliation{Department of Chemistry, National Taiwan University, Taipei 10617, Taiwan}
\author{Pei-Hsi Lee}
\affiliation{Graduate Institute of Photonics and Optoelectronics and Department of Electrical Engineering, National Taiwan University, Taipei 10617, Taiwan}
\author{Hung-Yi Lin}
\affiliation{Graduate Institute of Photonics and Optoelectronics and Department of Electrical Engineering, National Taiwan University, Taipei 10617, Taiwan}
\author{Bo-Yen Lin}
\author{Man-kit Leung}\affiliation{Department of Opto-Electronic Engineering, National Dong Hwa University, Shoufeng, Hualien 974301, Taiwan}
\author{Tien-Lung Chiu}\affiliation{Graduate Institute of Photonics and Optoelectronics and Department of Electrical Engineering,\\ National Taiwan University, Taipei 10617, Taiwan}
\author{Jiun-Haw Lee}\affiliation{Graduate Institute of Photonics and Optoelectronics and Department of Electrical Engineering,\\ National Taiwan University, Taipei 10617, Taiwan}
\author{Richard H. Friend}
\affiliation{Cavendish Laboratory, University of Cambridge, Cambridge CB3 0HE, UK}

\author{Yuh-Renn Wu}
\email{yrwu@ntu.edu.tw}
\affiliation{Graduate Institute of Photonics and Optoelectronics and Department of Electrical Engineering,\\ National Taiwan University, Taipei 10617, Taiwan}

\date{\today}



\begin{abstract}
In this study, we develop a steady state and time-dependent exciton diffusion model including singlet and triplet excitons coupled with a modified Poisson and drift-diffusion solver to explain the mechanism of hyper triplet--triplet fusion (TTF) organic light-emitting diodes (OLEDs). Using this modified simulator, we demonstrate various characteristics of OLEDs, including the J-V curve, internal quantum efficiency, transient spectrum, and electric profile. This solver can also be used to explain the mechanism of hyper-TTF-OLEDs and analyze the loss from different exciton mechanisms. Furthermore, we perform additional optimization of hyper-TTF-OLEDs that increases the internal quantum efficiency by approximately 33\% (from 29\% to 40\%).
\end{abstract}
\maketitle


\section{Introduction}
Since the first organic light-emitting diodes (OLEDs) were fabricated by Tang and VanSlyke  \cite{tang1987organic}, they have attracted the attention of researchers and consumers because of their ability to emit light that is very similar to natural light. Recently, several OLED-based devices (e.g., smartphones, wearable devices, and AR/VR/TV applications) have been developed as consumer electronics \cite{salehi2019recent,display1,LED1,chen2018liquid,xiong2021augmented}. However, the efficiency of OLEDs is still lower than that of traditional solid-state lighting devices, especially blue OLEDs \cite{bender2015solid,huang2020mini,hsiang2021prospects}. The reason is that the emitting mechanism of solid-state light devices is different from that of OLEDs. For solid-state lighting devices, the electron and hole directly recombine as a photon through fluorescent processes, whereas the electron and hole form as singlet and triplet excitons before emitting a photon in OLEDs. The long lifetime of triplet photons means that excitons might experience several decay processes before the exciton transforms into a photon.

First-generation OLED devices (fluorescent OLEDs) had a disadvantage in terms of a lower external quantum efficiency (EQE) because they only used a singlet exciton to generate fluorescence. However, according to the spin selection rule, 25\% singlet and 75\% triplet excitons are generated from electron and hole recombination \cite{sintri1,sintri2,sintri3}, so the theoretical maximum internal quantum efficiency (IQE) is only 25\%. Second-generation phosphorescent (Ph) OLED devices implement a triplet exciton mechanism to generate phosphorescence \cite{baldo1998highly}. Although PhOLED devices achieve higher performance, they contain heavy metal compounds (iridium-based organic compounds), which increases costs \cite{baldo1998highly,baldo1999very,adachi2001nearly}.

\begin{figure}[!bpht]
\resizebox{80mm}{!}{\includegraphics{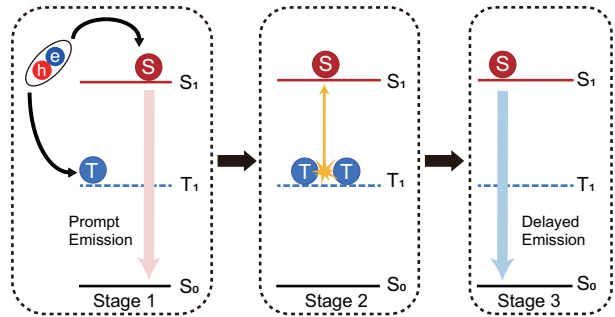}}
\centering
\caption{Schematic illustration of mechanism of TTF-OLED emission process.}
\label{fig1}
\end{figure}

Hence, considerable research has been dedicated to finding the next generation of OLEDs. The candidates for these next-generation OLEDs use both singlet and triplet excitons as much as possible. Two potential candidates are thermally activated delayed fluorescence (TADF) and triplet--triplet fusion (TTF) OLEDs. The former can achieve a theoretical IQE of 100\% (25\% + 75\%), whereas the latter has a theoretical IQE of 62.5\% (25\% + 75\%/2) if higher triplet and quintet states are inaccessible \cite{kuik2014g, uoyama2012highly,cui2017controlling,lin2018probe,lin2016sky,kondakov2009triplet,gao2021application,kondakov2015triplet,di2017efficient}. However, PhOLEDs and TADF-OLEDs suffer from the same issues as blue-OLEDs, which have worse EQE. The device lifetime of blue-OLEDs is shorter than that of red- and green-OLEDs, because they rely on high-energy and longer-lifetime triplet excitons to achieve emission, resulting in material degradation \cite{lee2019blue}. Hence, TTF-OLEDs are regarded as having the greatest potential as next-generation commercialized blue-OLEDs \cite{lee2019blue}. Figure \ref{fig1} shows the process of TTF-OLEDs. This kind of OLED utilizes a singlet exciton to emit prompt fluorescence (PF) in the first stage. The next stage involves the singlet undergoing the triplet--triplet annihilation-upconversion (TTA-UC) process, which generates delayed fluorescence (DF).

However, there is still a concern about TTF-OLEDs. The main reason is the difficulty of utilizing both singlet and triplet excitons to generate light, because the triplet exciton (which has a longer lifetime) is likely to be annihilated or quenched by polaron and singlet excitons in the emitting layer (EML). Hence, the reported performance of TTF-OLEDs is lower than the theoretical EQE of 12.5\% (assuming a light extraction efficiency of 20\%), while the delay ratio (the ratio of PF to DF) is usually less than 25\%. Hence, TTF-OLEDs cannot achieve efficient usage of the triplet excitons. To overcome this problem, many groups have proposed a new hyper-structure for TTF-OLEDs \cite{chen2022hyper,tasaki2022realization}. The primary concept is the implementation of a triplet tank layer (TTL) to separate the recombination zone, singlet excitons, and triplet excitons, thus avoiding triplet--polaron quenching (TPQ) and triplet--singlet annihilation (TSA) \cite{TPQ1, TPQ2}. Using hyper-TTF-OLEDs, Lee et al. achieved a recorded high EQE of 15.4\% for blue TTF-OLEDs and a high delay ratio of 33\% \cite{chen2022hyper}. To clarify the mechanism of hyper-TTF-OLEDs, quantitative numerical modeling is needed \cite{van2022numerical}. However, current commercial TCAD software only supports singlet or triplet exciton behavior. Next-generation OLED devices cannot be simulated because it is not possible to handle the exciton coupling between singlets and triplets. Hence, in this study, we develop a complete exciton diffusion model considering both singlet and triplet excitons and their interactions (TSA, TTA, and TTF) as a means of demonstrating the performance of TTF-OLEDs.

\vspace*{\fill}
\vspace*{\fill}

\vspace*{\fill}
\begin{figure*}[!bpht]
\resizebox{140mm}{!}{\includegraphics{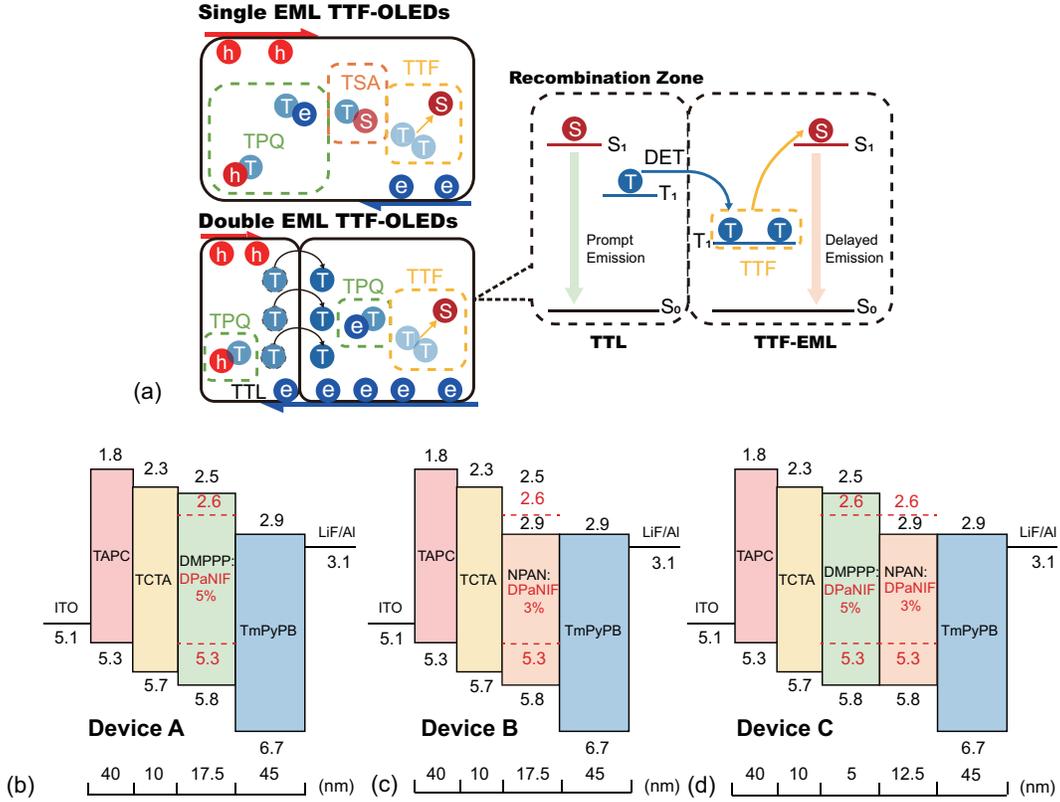}}
\centering
\caption{(\textbf{a}) Schematic showing the mechanism of typical and hyper-TTF-OLEDs. (\textbf{b})--(\textbf{d}) Energy levels and structures of devices A, B, and C, respectively.}
\label{fig2}
\end{figure*} 
\vspace*{\fill}

\vspace*{\fill}

\section{Methodology}
The simulation program developed in this study is based on a modified Poisson--drift-diffusion (DD) solver, which can be utilized for organic-based devices with a Gaussian density of state and a field-dependent mobility model. We develop an exciton diffusion model that considers most exciton behaviors and the interaction of singlet and triplet excitons. This exciton diffusion model includes steady state and time-dependent modes. The steady state solver can be used to calculate the quantum efficiency of the device, while the time-dependent mode can be used to demonstrate the transient spectrum of the device and extract the parameters of singlet and triplet excitons through calibrations against experimental data. The experimental data are taken from a study by Lee et al. \cite{chen2022hyper}.





\subsection{Poisson and drift-diffusion solver}
A modified 1D Poisson-DD solver is applied to simulate the characteristics of TTF-OLED devices. This solver was developed by our laboratory; for more details, see Ref.~\cite{yrwu}. This solver is based on the Poisson, DD, and continuity equations:
\begin{equation}
\nabla_{r}\cdot(\varepsilon\nabla_{r}V(r))=q(n_{free}-p_{free}+N^{-}_{a}-N^{+}_d+\cdots),
\label{eq-poisson}
\end{equation}
\begin{equation}
J_{n}=-q\mu_{n}n_{free}(r)\nabla_{r}V(r)+qD_{n}\nabla_{r}n_{free}(r),
\label{eq-Jn}
\end{equation}
\begin{equation}
J_{p}=-q\mu_{p}p_{free}(r)\nabla_{r}V(r)-qD_{p}\nabla_{r}p_{free}(r),
\label{eq-Jp}
\end{equation}
\begin{equation}
\nabla_{r}(J_{n,p})=q(R-G).
\label{eq-con}
\end{equation}

Equation (\ref{eq-poisson}) is the Poisson equation, where $\rm \varepsilon$ is the dielectric constant at different device positions, $V$ is the potential energy of the device, and $n_{free}$ and $p_{free}$ are the electron and hole densities, respectively. In the DD equations [Eqs.~(\ref{eq-Jn}) and (\ref{eq-Jp})], $J_{n}$ and $J_{p}$ are the current densities of electrons and holes, respectively; $\mu_{n}$ and $\mu_{p}$ are the mobilities of electrons and holes, respectively; $D_{n}$ and $D_{p}$ are the diffusion coefficients of electrons and holes, respectively; and $R$ is the recombination rate shown in the following:

\begin{equation}
R=SRH + Bnp + C_0(n^2p+p^2n),
\label{eq-R}
\end{equation}
where B is the radiative-recombination rate, and $\rm C_0$ is the coefficient of Auger recombination. However, generally, there is merely no Auger recombination in OLEDs. Moreover, SRH is Shockley-Read-Hall recombination, which is shown in the following:
\begin{equation}
SRH=\frac{np-n_i^2}{\tau_{n0}(p+n_i) + \tau_{p0}(n+n_i)},
\label{eq-SRH}
\end{equation}

where $n$ and $p$ are electron and hole density, respectively; $\tau_n$ is electron carrier lifetime, and $\tau_p$ is hole carrier lifetime, and $n_i$ is intrinsic carrier density.

For organic materials, the carrier lifetime in Poisson-DD solver are set as quite long because most electron-hole pair would become exciton before they decay. Hence, the main non-radiative recombination mechanism is non-radiative exciton decay in the exciton diffusion model.

\subsubsection{Gaussian density of state}
The energy bands of organic materials are the highest occupied molecular orbital (HOMO) and the lowest unoccupied molecular orbital (LUMO). Although these bands are similar to valance bands and conduction bands in semiconductors, organic materials exhibit different properties from semiconductors. The major difference is that the band edges are not abrupt. Hence, there are some tail states in the bandgap, especially around LUMO and HOMO, which means the density of the state is a disordered distribution. A previous study has shown that this disordered distribution can be regarded as a Gaussian density of state \cite{gaussian1,gaussian2,gaussian3}, which can be utilized to describe carrier transport in organic materials. Hence, the following Gaussian density of state is implemented to describe the tail states:
\begin{equation}
N_{tail}(E)=N_{t}\frac{1}{\sigma\sqrt{2\pi}}exp\left[-\frac{(E-E_t)^2}{2\sigma^{2}}\right],
\label{eq10}
\end{equation}
where $N_t$ is the total density of state, $\sigma$ is the broadening factor of the Gaussian shape, and $E_{t}$ is the difference between the center of the Gaussian density of state and the LUMO or HOMO.

When a Gaussian density of state is applied, the carrier density can be expressed as 
\begin{equation}
n, p=\int_{-\infty}^{\infty} N_{tail}(E)f_{e,h}(E) \, dE;
\label{eq11}
\end{equation}
otherwise, the carrier density is expressed as
\begin{equation}
n=\int_{-\infty}^{\infty} \frac{1}{2\pi^2} \Big(\frac{2m^*_{e}}{\hbar^2}\Big)^{\frac{3}{2}} \sqrt{E-E_c}~f_{e}(E) \, dE,
\label{eq12}
\end{equation}
and

\begin{equation}
p=\int_{-\infty}^{\infty} \frac{1}{2\pi^2} \Big(\frac{2m^*_{h}}{\hbar^2}\Big)^{\frac{3}{2}} \sqrt{E_v-E}~f_{h}(E) \, dE,
\label{eq13}
\end{equation}
where $m_{e,h}$ is the effective mass of the electron or hole, respectively; $E_{c,v}$ is the conduction or valance band, respectively; and $f_{e,h}$ is the Fermi--Dirac distribution function for the electron or hole, respectively.

\subsubsection{Field-dependent mobility model}
Many studies have shown that carrier transport in organic materials is a hopping process. The mobility can be accelerated by applying electric fields. Hence, the mobility behavior is described with the Poole--Frenkel model \cite{mu1,mu2,mu3,mu4}. The Poole--Frenkel field-dependent mobility model can be written as follows:

\begin{equation}
\mu_{e,h}=\mu_{0_{e,h}} \times exp\left(\beta_{e,h}\sqrt{\vec{|F|}}\right),
\label{eq-mobility}
\end{equation}
where $\mu_{e,h}$ is the carrier mobility for electrons and holes, respectively; $\mu_{0_{e,h}}$ is the zero-field mobility for electrons and holes, respectively; $\beta_{e,h}$ is the field activation parameter for electrons and holes, respectively; and $\vec{F}$ is the electric field.

\subsection{Exciton diffusion solver}
To demonstrate the performance of TTF-OLEDs, both singlet and triplet excitons must be considered because the mechanism of TTF-OLEDs involves (1) prompt emission from the singlet radiative process and (2) delayed emission from TTF to the singlet radiative process. In this work, an exciton diffusion solver that considers the interaction of both singlet and triplet excitons is developed. Most exciton behaviors are considered in this exciton solver, including exciton diffusion, exciton radiative and non-radiative decay, electron- and hole-induced TPQ, TSA, TTA, and TTF. Additionally, the energy transfer rate of the excitons is introduced to the exciton diffusion equation at the interface of the heterojunction. Details can be found in our previous study \cite{huang2020analysis}. The singlet and triplet exciton diffusion equations are written as follows:

\begin{equation}
\frac{dn^S_{ex}}{dt}= D^S_{ex} \nabla_r^2 n^S_{ex}-\left(k^S_r+k^S_{nr}+k^S_{e} n+k^S_{h} p+k_{TS}n^T_{ex}\right)n^S_{ex}+\frac{1}{2}\gamma_{TS} {n^{T}_{ex}}^2+ G^S_{ex},
\label{eq-exciton}
\end{equation}

\begin{equation}
\frac{dn^T_{ex}}{dt}= D^T_{ex} \nabla_r^2 n^T_{ex}-\left(k^T_r+k^T_{nr}+k^T_{e} n+k^T_{h} p+k_{TS}n^S_{ex}\right)n^T_{ex}-(\gamma_{TS}+\gamma_{TT}) {n^{T}_{ex}}^2+ G^T_{ex},
\label{eq-exciton2}
\end{equation}
where $n_{ex}$ is the exciton density distribution, $D_{ex}$ is the exciton diffusion coefficient, $k_r$ and $k_{nr}$ are the radiative and non-radiative exciton constants, respectively, $k_{e,h}$ is the TPQ rate constant for electrons and holes, respectively, and $G_{ex}$ is the initial exciton density distribution. The generations of singlet and triplet excitons constitute 1/4 and 3/4 of the radiative recombination rate (R) from the Poisson-DD solver, respectively. The superscripts S and T represent singlet and triplet excitons, respectively. $\gamma_{TS}$ is the TSA coefficient, $\gamma_{TT}$ is the TTA coefficient, and $n$, $p$ are the carrier densities of electrons and holes, respectively.

\begin{figure}[!bpht]
\resizebox{150mm}{!}{\includegraphics{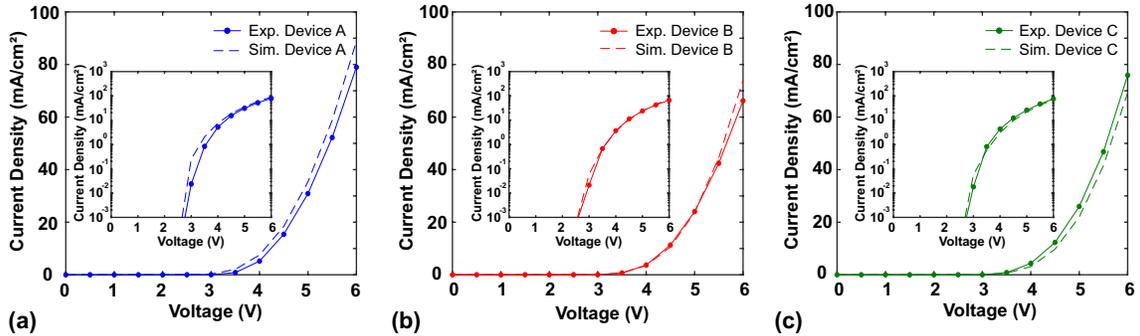}}
\centering
\caption{(\textbf{a})-(\textbf{c}) The characteristic of J-V curves for devices A, B, and C, respectively.}
\label{newIV}
\end{figure} 
\vspace*{\fill}

\vspace*{\fill}

\begin{figure}[!bpht]
\resizebox{120mm}{!}{\includegraphics{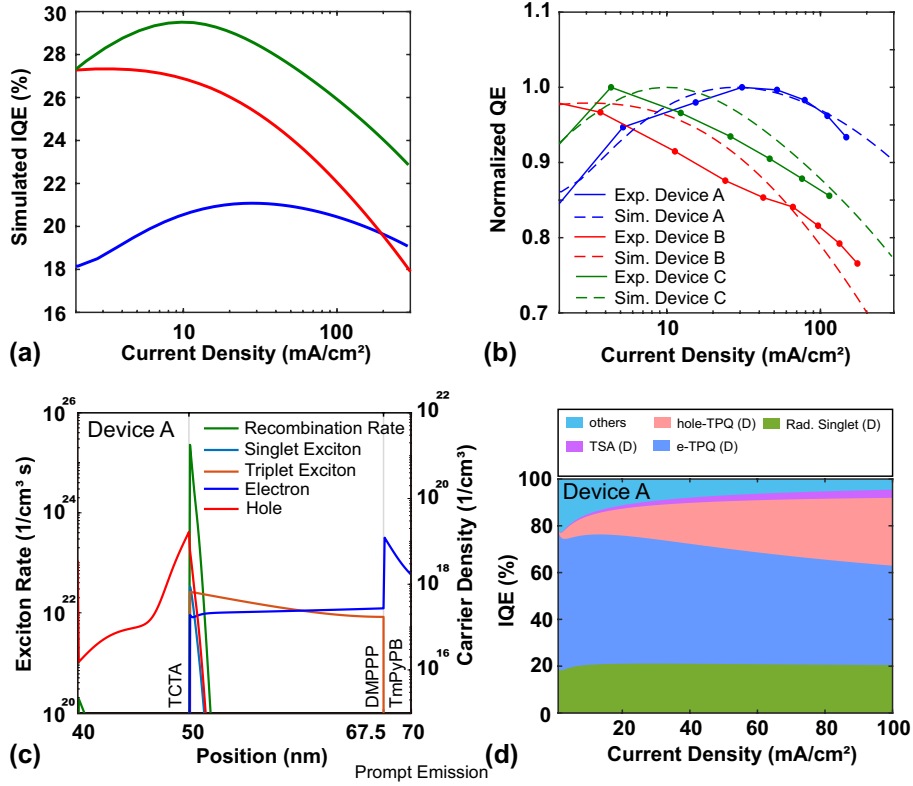}}
\centering
\caption{(\textbf{a}) Simulated IQE curves versus current density for devices A, B, and C. (\textbf{b}) The simulated and experimental normalized QE versus current density for devices A, B, and C. (\textbf{c}) Recombination rate, singlet excitons, triplet excitons, electron density, and hole density for device A. (\textbf{d}) Loss results from radiative singlet exciton, electron-induced TPQ, hole-induced TPQ, and TSA for device A.}
\label{fig3}
\end{figure} 
\vspace*{\fill}

\vspace*{\fill}
\begin{table*}[!bpht]
\small
\caption{Parameters of the density of states and mobility used in this work.}
\centering
\begin{tabular}{lccccccccccc}
\hline\hline
Material &$N_{t,e}$ & $E_{t,e}$ & $\sigma_{t,e}$ & $N_{t,h}$& $E_{t,h}$ & $\sigma_{t,h}$ & $\rm\mu_{0,e}$ & $\rm\beta{e}$ & $\rm\mu_{0,h}$& $\rm\beta_{h}$ \\
& (cm$^{-3}$) &(eV)&(eV)& (cm$^{-3}$) &(eV)&(eV)&(cm$\rm^{2}$/Vs) &(cm/V)$\rm^{0.5}$ & (cm$\rm^{2}$/Vs) & (cm/V)$\rm ^{0.5}$ \\
\hline
TAPC & 1.0$ \times 10^{19}$ & 1.80 & 0.06 & 1.0$ \times 10^{19}$ & 5.10 & 0.12 & 1.0$\times 10^{-8}$ & 0.0070 & 8.7$\times 10^{-4}$ \cite{huh2013high}& 0.0018 \cite{huh2013high}\\
TCTA & 1.0$ \times 10^{19}$ & 2.00 & 0.10 & 1.0$ \times 10^{17}$ & 5.90 & 0.08 & 2.2$\times 10^{-6}$ & 0.0070 & 5.5$\times 10^{-5}$ \cite{li2012effects}& 0.0024 \cite{li2012effects}\\
TmPyPB & 1.0$ \times 10^{19}$ & 2.80 & 0.12 & 1.0$ \times 10^{19}$ & 6.70 & 0.12 & 3.2$\times 10^{-4}$ \cite{su2008pyridine}& 0.0014 \cite{su2008pyridine}& 2.4$\times 10^{-8}$ & 0.0048\\
DMPPP: DPaNIF 5\% & 1.0$ \times 10^{18}$ & 2.70 & 0.10 & 1.0$ \times 10^{18}$ & 5.75 & 0.08 & 8.0$\times 10^{-7}$ & 0.0007 & 5.0$\times 10^{-9}$ & 0.0010\\
NPAN: DPaNIF 3\% & 1.0$ \times 10^{18}$ & 2.70 & 0.10 & 1.0$ \times 10^{18}$ & 5.75 & 0.08 & 5.0$\times 10^{-9}$ & 0.00375& 3.0$\times 10^{-8}$ & 0.0007\\
\hline \hline
\end{tabular}
\label{tab1}
\end{table*}
\vspace*{\fill}

\vspace*{\fill}
\begin{table*}[!bpht]
\scriptsize
\caption{Parameters of exciton diffusion model used in this work.}
\centering
\begin{tabular}{ccccccccccc}
\hline\hline
Material &$D^S_{ex}$ & $D^T_{ex}$ & $k_r^S$ & $k_{nr}^S$ & $k_r^T$ & $k_{nr}^T$ & $k_{e,h}^T$ & $k_{TS}$ & $\gamma_{TS}$& $\gamma_{TT}$ \\
& (cm$^{-2}$/s) &(cm$^{-2}$/s)&(1/s)& (1/s) &(1/s)& (1/s)&(cm$\rm^{3}$/s) &(cm$\rm^{3}$/s)&(cm$\rm^{3}$/s) &(cm$\rm^{3}$/s) \\
\hline
TAPC &1.0$ \times 10^{-11}$ & 1.0$ \times 10^{-11}$ & 0 & 4.0$ \times 10^{8}$ & 0 & 4.0$ \times 10^{6}$ & 0 & 0 & 0 & 0\\
TCTA &1.0$ \times 10^{-11}$ & 1.0$ \times 10^{-11}$ & 0 & 4.0$ \times 10^{8}$ & 0 & 4.0$ \times 10^{6}$ & 0 & 0 & 0 & 0\\
TmPyPB &1.0$ \times 10^{-11}$ & 1.0$ \times 10^{-11}$ & 0 & 4.0$ \times 10^{8}$ & 0 & 4.0$ \times 10^{6}$ & 0 & 0 & 0 & 0\\
DMPPP: DPaNIF 5\% & 1.0$ \times 10^{-7}$ & 1.0$ \times 10^{-7}$ & 4.0$ \times 10^{8}$ & 4.0$ \times 10^{6}$ &0 & 1.0$ \times 10^{4}$ & 4.0$ \times 10^{-13}$ & 2.5$ \times 10^{-11}$ & 0 & 1.0$ \times 10^{-15}$ \\
NPAN: DPaNIF 3\% &5.0$ \times 10^{-9}$ & 5.0$ \times 10^{-9}$ & 3.0$ \times 10^{8}$ & 9.0$ \times 10^{6}$ &0 & 1.0$ \times 10^{4}$ & 3.0$ \times 10^{-13}$ & 8.0$ \times 10^{-11}$ & 1.0$ \times 10^{-13}$ & 1.0$ \times 10^{-15}$\\
\hline \hline
\end{tabular}
\label{tab2}
\end{table*}
\vspace*{\fill}

\section{Results and Discussion}
Typical TTF-OLEDs only use a single EML, as shown in Fig.~\ref{fig2}(a). This structure would cause all of the carrier recombination zone, singlet emission zone, and TTF processing region to be generated in the same layer \cite{di2017efficient,lee2019blue}. Hence, this strategy results in significant TPQ and TSA. Moreover, the TTA-UC processing rate is usually slower than that of TSA and TPQ, so most triplet excitons are eliminated by polarons and singlet excitons before transforming into singlet excitons through the UC mechanism \cite{murawski2013efficiency,lee2019blue}. Therefore, the highest reported EQE of TTF-OLEDs is less than 12\%, which is close to the minimal theoretical value of 12.5\% (with light extraction efficiency of 20\%) \cite{chen2016superior,chou2014efficient,wang2011new}.

As mentioned previously, Lee et al. recently proposed a hyper-TTF structure to overcome this problem \cite{chen2022hyper}. Figure \ref{fig2}(a) shows the typical structure of hyper-TTF-OLEDs. The primary strategy of hyper-TTF-OLEDs is to use a TTL to separate the singlet and triplet excitons, which prevents the efficiency loss resulting from TSA and TPQ. Hence, three key points are required: (1) the TTL should be implemented by a non-TTA-UC organic emitter; (2) the electron--hole recombination zone should be concentrated in the region of TTL; and (3) the TTL's exciton diffusion coefficient should be as high as possible. Additionally, (4) a high photoluminescence quantum yield (PLQY) is needed because we would like most triplet exciton transfers to another organic emitter to occur through the TTA-UC process before TSA, as this will achieve higher performance.

To analyze the mechanism of hyper TTF-OLEDs, three different devices with different EML structures are analyzed by our solver: (i) device A is a non-TTF-OLED (DMPPP: 5\% DPaNIF), (ii) device B is a TTF-OLED (NPAN: 3\% DPaNIF), and (iii) device C is a hyper-TTF-OLED (double emitter, DMPPP: 5\% DPaNIF and NPAN: 3\% DPaNIF). These devices were considered in the previous study of Lee et al. \cite{chen2022hyper} To exclude the effects of carrier injection by these layers, the cathode (ITO), hole transport layer (TAPC, 40 nm), electron blocking layer (TCTA, 40 nm), electron transport layer (TmPyPB, 45 nm), electron injection layer (LiF, 1 nm), and anode (Al, 100 nm) are fixed in these devices. Figures \ref{fig2}(b)--\ref{fig2}(d) show that the EMLs are different (DMPPP: 5\% DPaNIF 17.5 nm, NPAN: 3\% DPaNIF 17.5 nm, DMPPP: 5\% DPaNIF 5 nm, and NPAN: 3\% DPaNIF 12.5 nm). Detailed experimental measurements, including the TREL spectra, efficiency versus current density, J-V curve, and PLQY results for these devices, can be found in Prof. Lee et al.'s previous work \cite{chen2022hyper}.


The carrier and exciton distributions and behavior are difficult to observe or measure experimentally. Hence, the simulated results are presented in the following section. These results can be used to explain the mechanism of hyper-TTF-OLEDs and implement further optimization.

\subsection{Device modeling of hyper-TTF-OLEDs}
Before analyzing the efficiency and mechanism of these hyper-TTF-OLEDs, the J-V curves, TREL spectra, and EQE characteristic curves of devices A, B, and C were obtained by our solver. The modeling parameters used in this work are listed in Tables \ref{tab1} and \ref{tab2}. For the parameters in exciton diffusion model, the radiative and non-radiative singlet exciton decay constants are from PLQY and transient EL spectrum. The details are shown in figures S1 and S2 and table S1 in Supporting Information. Figures \ref{newIV}(a)-(c) show the J-V characteristic curves of the three different devices including simulation and experiment. The J-V characteristics of each device are similar because the injection condition and electrical impedance are the same in all devices. Figure \ref{fig3}(a) shows that there are significant variations in the trends of EQE versus current density for the three devices. The simulated IQE can not compare with experimental results directly because the light extraction efficiency of different devices is not the same. Hence, the normalized QE of simulation and experiment is utilized to exclude the effect of different light extraction efficiencies. Figure \ref{fig3}(b) shows that the simulated and experimental QE for device A, B, and C. And the raw data of experimental EQE for all devices has been shown in figure S3 in the Supporting Informatoin.

Overall, device A exhibits lower EQE performance, although its efficiency decay with increasing current density is the slowest of all cases. The reason can be explained by the carrier and exciton distributions of device A. Figure \ref{fig3}(c) shows the carrier and exciton distributions at a voltage of 6.0 V, and the key points causing the lower efficiency can be identified. There is an inadequate carrier injection between the EML and ETL because the difference in energy between these two materials is 0.3 eV. Moreover, device A only utilizes singlet excitons to generate PF because triplet exciton cannot generate TTA-UC to transform into singlet excitons in the DMPPP layer. Hence, the maximum theoretical IQE is only 25\%. However, due to the efficient Dexter energy transfer of DMPPP, Fig.~\ref{fig3}(c) shows that the triplet excitons diffuse well in the EML and their distribution is smooth, which results in a lower TSA in device A. Figure \ref{fig3}(d) shows that TSA is infrequent, resulting in an efficiency loss for device A. Many studies have reported that TSA is the main reason for the efficiency roll-off in fluorescent OLEDs. However, the high $D_{ex}^{T}$ value in DMPPP means that this drop-off effect at higher current densities is not significant. 

\vspace*{\fill}
\begin{figure}[!bpht]
\resizebox{150mm}{!}{\includegraphics{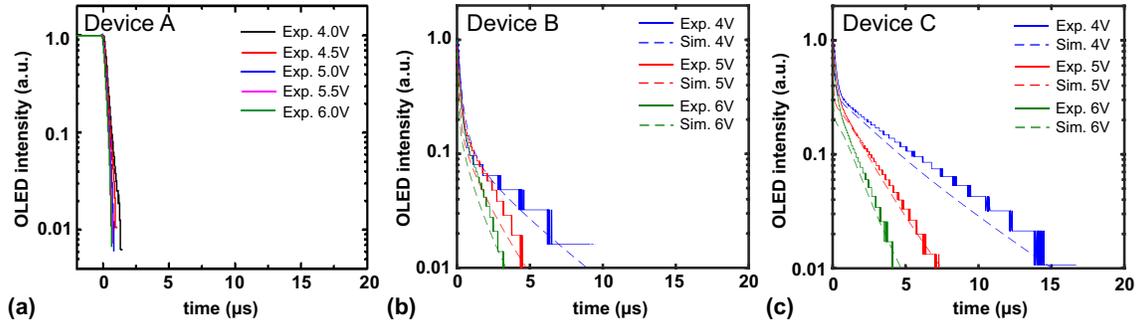}}
\centering
\caption{(\textbf{a})--(\textbf{c}) TREL spectra for devices A, B, and C, respectively. The solid and dashed lines show the experimental and simulated results, respectively.}
\label{TREL}
\end{figure} 
\vspace*{\fill}

\begin{figure}[!bpht]
\resizebox{120mm}{!}{\includegraphics{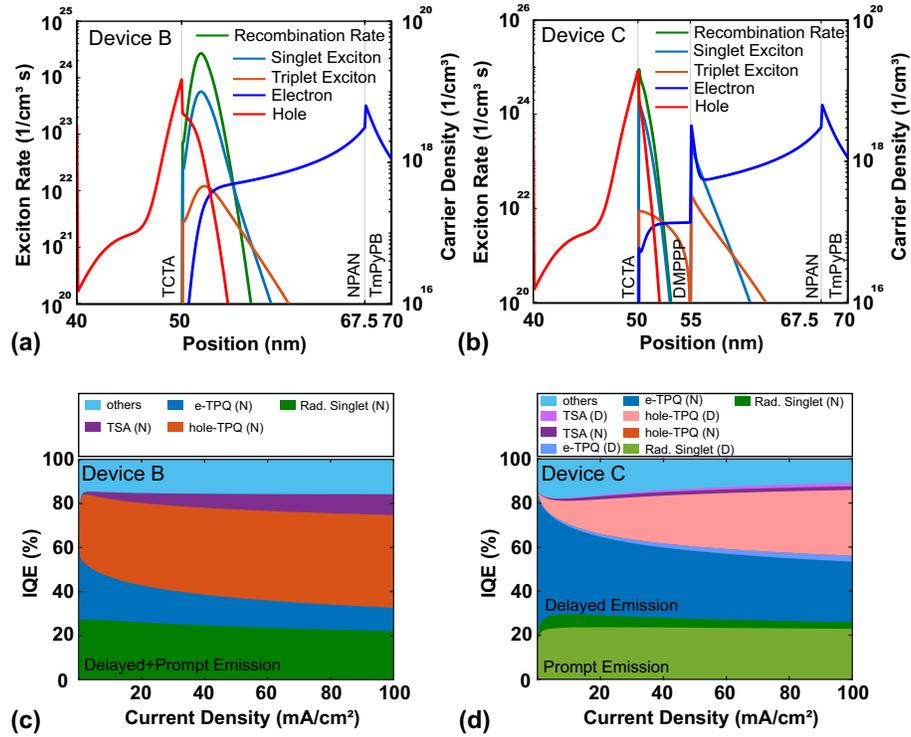}}
\centering
\caption{(\textbf{a}), (\textbf{b}) Recombination rate, singlet excitons, triplet excitons, electron density, and hole density for devices B and C, respectively. (\textbf{c})--(\textbf{d}) Loss resulting from radiative singlet excitons, electron-induced TPQ, hole-induced TPQ, and TSA for devices B and C, respectively.}
\label{fig4}
\end{figure}

Figure \ref{TREL}(a) shows that device A only emits PF, while Figs.~\ref{TREL}(b) and \ref{TREL}(c) show that devices B and C generate both PF and DF. Even though both device B and device C can utilize prompt singlets and delayed singlets from the TTA-UC mechanism to generate PF and DF, the delay ratio of device C is better than that of device B. Hence, device C exhibits better overall performance compared with device B. Additionally, device C has a slower efficiency roll-off. Figure \ref{fig4}(a) shows the exciton and carrier distributions in device B. It can be observed that device B is a typical TTF-OLED, where the recombination zone, singlet and triplet exciton distributions, and electron and hole density distributions are in the same region. Hence, the efficiency of device B at the lower current density is higher than that of device A. At the higher current density, device B has greater efficiency than device A because of the stronger TPQ and TSA. Figure \ref{fig4}(c) shows that the TPQ and TSA loss is severe in device B, resulting in a significant efficiency roll-off at the higher current density.

Figure \ref{fig4}(b) shows the carrier and exciton distributions at a voltage of 6.0 V for device C (hyper-TTF-OLED). This figure shows that the recombination zone, hole density, and prompt singlet excitons are in the DMPPP layer, while the electron density and most triplet excitons are in the NPAN layer. Although excitons are initially only generated in the DMPPP layer, singlet and triplet excitons can still be observed in the NPAN layer. This is because triplet excitons can be transferred from the DMPPP and singlet excitons are transformed through the TTA-UC mechanism in the NPAN layer. Hence, the recombination zone, singlet excitons, and triplet excitons can be separated by introducing a TTL in hyper-TTF-OLEDs. Figure \ref{fig4}(d) shows the lower loss from the TSA mechanism in device C; recall that TSA is quite severe at the higher current density in device B. However, Figs.~\ref{fig4}(c) and \ref{fig4}(d) still show that device C has a significant loss resulting from hole- and electron-induced TPQ. The reason is the numerous electrons accumulating at the right side of the interface between the DMPPP and NPAN layers, and the holes concentrated in the DMPPP layer. Hence, the performance of hyper-TTF-OLEDs can be further optimized by improving the carrier injection and distribution to avoid or decrease the loss resulting from electron- and hole-induced TPQ. The next section presents numerical modeling results that demonstrate how to reduce the electron- and hole-induced TPQ effect on the performance of hyper-TTF-OLEDs.

\vspace*{\fill}
\vspace*{\fill}

\subsection{Optimization of hyper-TTF-OLEDs} 
The modeling results show that the hyper-TTF-OLED structure can significantly reduce the TSA effect and decrease the electron- and hole-induced TPQ effects. The addition of a TTL in hyper-TTF-OLEDs separates the singlet and triplet excitons into different regions, resulting in fewer TSA events. Although hyper-TTF-OLEDs can separate the delayed emission layer (NPAN) and hole carrier into different layers, as shown in Fig.~\ref{fig4}(c), there are still both electron- and hole-induced TPQs in hyper-TTF-OLEDs.

\vspace*{\fill}
\begin{figure}[!bpht]
\resizebox{120mm}{!}{\includegraphics{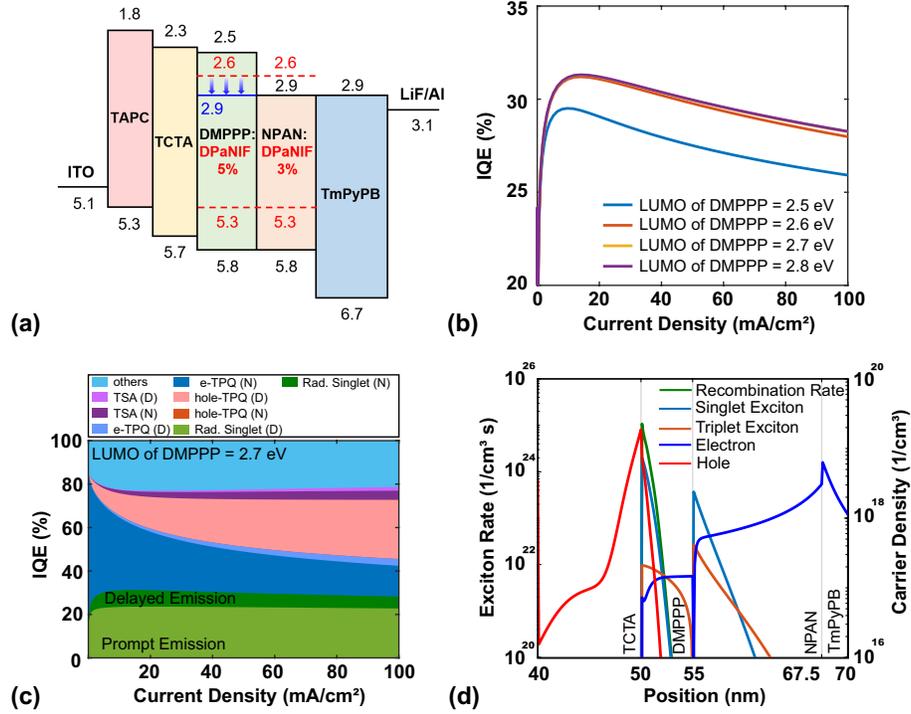}}
\centering
\caption{(\textbf{a}) Optimized structure of hyper-TTF-OLEDs. (\textbf{b}) IQE versus current density for cases with different LUMO in the DMPPP layer. (\textbf{c}) Losses resulting from radiative singlet excitons, TPQ, and TSA in different layers for the device with a DMPPP layer LUMO of 2.7 eV. (\textbf{d}) Recombination rate, singlet excitons, triplet excitons, electron density, and hole density for the device with a DMPPP layer LUMO of 2.7 eV.}
\label{fig5}
\end{figure}
\vspace*{\fill}

\subsubsection{Optimization of hyper-TTF-OLEDs---LUMO of TTL}
Both electron- and hole-induced TPQs are hard to avoid because both singlet and triplet excitons are transformed from electron--hole pairs and are initially generated in the same layer. However, reducing the TPQ effect is feasible by choosing an ideal TTL made from a material with suitable properties in the hyper-TTF-OLED device. Figure \ref{fig4}(b) shows that many electrons accumulate at the right side of the interface between the DMPPP layer and the NPAN layer, which is the main reason for the severe electron-induced TPQ effect in the NPAN layer [see Fig.~\ref{fig4}(d)]. Figure \ref{fig2}(d) shows that, in the band alignment of device C, there is an energy barrier between the DMPPP and NPAN layers (0.3 eV), and the electron mobility of the NPAN layer is only $\rm 5\times10^{-9}$ $\rm cm^2/Vs$. Hence, there is a serious electron accumulation in the NPAN layer. 

Figure \ref{fig5}(a) shows the optimized structure of hyper-TTF-OLEDs as the LUMO of the TTL changes from 2.5 eV to 2.9 eV to reduce the accumulation of electrons at the interface between the DMPPP and NPAN layers. Figure \ref{fig5}(b) shows the IQE versus current density for cases of different LUMOs in the TTL. There is a noticeable efficiency improvement of around 6\%. However, the results for the cases from 2.7--2.9 eV are the same. Figure \ref{fig5}(c) shows that the main improvement comes from reducing the electron-induced TPQ in the NPAN layer. There are more triplet excitons, which are transferred to the DMPPP layer to induce the TTA-UC mechanism. This results in higher delayed singlet emissions and IQE than in device C. Figure \ref{fig5}(d) shows that the electron accumulation at the interface between the DMPPP and NPAN layers can be decreased by changing the LUMO of the TTL. However, the electron density in the NPAN layer is still higher than the electron density in the DMPPP layer. The reason is that the electron mobilities of DMPPP: 5\% DPaNIF and NPAN: 3\% DPaNIF differ by a factor of around 60. Therefore, the distribution of electron carriers is affected by the electron mobility. The influence of mobility will be discussed in the next section.
\vspace*{\fill}
\begin{figure}[!bpht]
\resizebox{120mm}{!}{\includegraphics{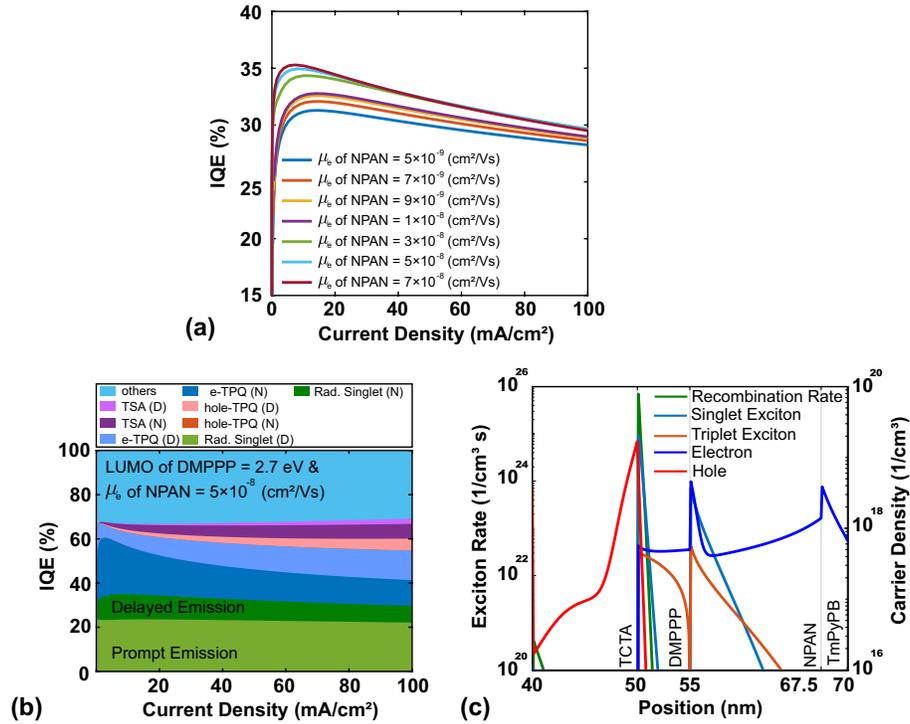}}
\centering
\caption{(\textbf{a}) Characteristics of J-V curve for cases with different electron mobilities in the NPAN layer. (\textbf{b}) Losses resulting from radiative singlet excitons, TPQ, and TSA for an electron mobility of $5\times10^{-8}$ $\rm cm^2/Vs$ in the NPAN layer. (\textbf{c}) Recombination rate, singlet excitons, triplet excitons, electron density, and hole density for an electron mobility of $\rm 5\times10^{-8}$ $\rm cm^2/Vs$ in the NPAN layer.}
\label{fig6}
\end{figure} 
\vspace*{\fill}

\subsubsection{Optimization of hyper-TTF-OLEDs---electron mobility of NPAN layer}
Because the electron mobility of the NPAN layer is lower than that of the DMPPP layer, even when the band alignment is optimized, there are still electrons accumulating in the NPAN layer, as shown in Fig.~\ref{fig5}(d). To overcome the electron-induced TPQ resulting from this accumulation over the whole NPAN layer, devices with NPAN layers of different electron mobilities are considered. Figure \ref{fig6}(a) shows the IQE versus current density for cases with different NPAN electron mobilities. The efficiency grows as the electron mobility increases. Figure \ref{fig6}(b) shows that both the electron- and hole-induced TPQ can be reduced by changing the mobility. Figure \ref{fig6}(c) illustrates that injecting more electron carriers into the DMPPP layer shifts the recombination zone to the left-hand side, which results in fewer hole carriers to quench the triplet excitons. Hence, the hole-induced TPQ decreases significantly when the electron mobility of the NPAN layer is modified. In addition, more triplet excitons are transferred to the DMPPP layer, resulting in the higher delayed singlet emissions in Fig.~\ref{fig6}(b). Additionally, more electrons are injected into the DMPPP layer, so the electron-induced TPQ is greater than $\rm 5\times10^{-8}$ $\rm cm^2/Vs$. In summary, the IQE of optimized hyper-TTF-OLEDs can be improved from 29\% to 35\%, which is effectively a 20\% improvement [from blue line to green line in Fig.~\ref{fig6}(a)].

\vspace*{\fill}
\begin{figure}[!bpht]
\resizebox{120mm}{!}{\includegraphics{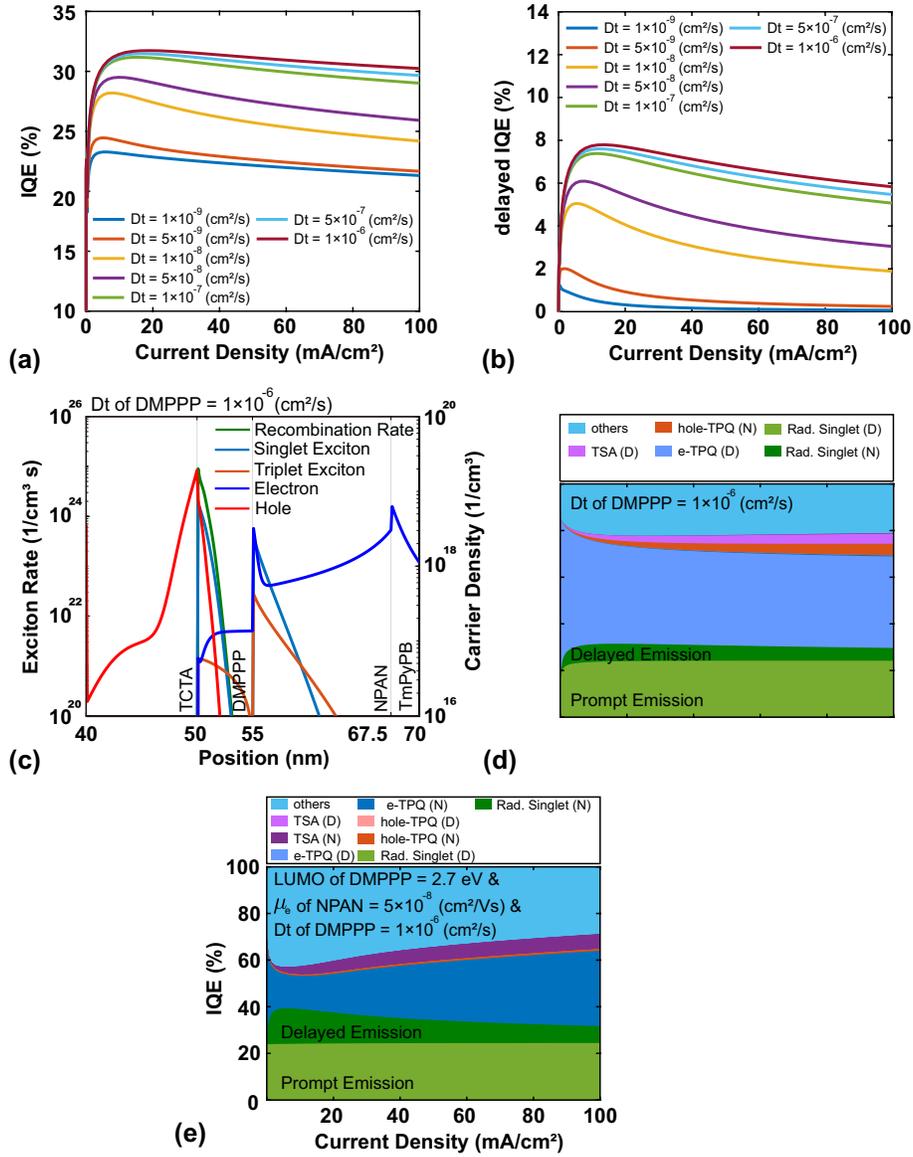}}
\centering
\caption{(\textbf{a}) IQE versus current density for cases with different diffusion coefficients of triplet excitons. (\textbf{b}) IQE of delayed emission versus current density for cases with different diffusion coefficient of triplet excitons. (\textbf{c}), (\textbf{d}) Recombination rate, singlet excitons, triplet excitons, electron density, and hole density and the losses resulting from radiative singlet excitons, electron-induced TPQ, hole-induced TPQ, and TSA for a triplet exciton diffusion coefficient of $\rm 1\times10^{-6}$ $\rm cm^2/s$ in the DMPPP layer. (\textbf{e}) Losses resulting from radiative singlet excitons, TPQ, and TSA in the optimized case.}
\label{figDrT}
\end{figure} 
\vspace*{\fill}

\subsubsection{Optimization of hyper-TTF-OLEDs---Dexter energy transfer of TTL}
Furthermore, the exciton properties of the TTL are crucial. As we have mentioned before, the function of the TTL is to separate singlet and triplet excitons into different layers. A TTL with efficient Dexter energy transfer is needed to achieve this goal. The value of the Dexter energy transfer affects the triplet diffusion coefficient in the exciton rate equation \cite{menke2014exciton,monguzzi2008upconversion}. This section presents results for different triplet exciton diffusion coefficients in the TTL, allowing us to discuss how the diffusion coefficient affects the performance of hyper-TTF-OLEDs. The simulated structure is shown in Fig.~\ref{fig2}(d) and the parameters are listed in Tables \ref{tab1} and \ref{tab2}.

Figure \ref{figDrT}(a) shows that the IQE of hyper-TTF-OLEDs is very sensitive to the value of the triplet exciton diffusion coefficient. Figure~\ref{figDrT}(b) shows there is no delayed singlet emission when the diffusion coefficient is less than $\rm 5\times10^{-8}$~$\rm cm^{2}/s$,  but the delayed singlet emission becomes saturated when the diffusion coefficient is greater than $\rm 5\times10^{-7}$ $\rm cm^2/s$. Figure \ref{figDrT}(c) indicates that this is because triplet excitons are still being annihilated by single excitons as the triplet exciton generation region is the DMPPP layer. Moreover, even though most triplets can transfer from the DMPPP layer to the NPAN layer, triplet excitons still experience the TPQ process when TTA-UC occurs. Hence, Fig.~\ref{figDrT}(d) shows that there is a significant loss resulting from electron-induced TPQ in the cases with higher diffusion coefficients. In summary, the TTL is a vital factor in hyper-TTF-OLEDs. For a typical TTL with a thickness of 5 nm, an organic material with a diffusion coefficient above $\rm 5\times10^{-8}$ $\rm cm^2/s$ should be used to ensure efficient delayed singlet emissions. Figure \ref{figDrT}(e) shows that the IQE of optimized hyper-TTF-OLEDs (with a DMPPP layer LUMO of 2.7 eV, an NPAN layer electron mobility of $\rm 5\times10^{-8}$ $\rm cm^2/Vs$, and a DMPPP layer triplet exciton diffusion coefficient of $\rm 1\times10^{-6}$ $\rm cm^2/s$) can be improved from 29\% to 40\%, an effective improvement of 33\%.
\section{Conclusion}
In this work, we developed a modified Poisson-DD solver and exciton diffusion solver that considers most exciton behavior (TPQ, TSA, TTA, and others) and the interaction between singlet and triplet excitons. Moreover, this model was used to analyze typical TTF- and hyper-TTF-OLEDs. TTF-OLEDs usually suffer from a severe TPQ and TSA losses because they usually implement a single EML. However, Lee et al. \cite{chen2022hyper} proposed hyper-TTF-OLEDs, which separate prompt and delayed singlets through the TTL, enabling a high EQE of 15.4\% in blue TTF-OLEDs. Using this simulation solver, we summarized some key points to achieve higher performance and delayed singlet emissions of hyper-TTF-OLEDs. The band alignment of the TTL and the delayed emission layer (NPAN layer in this work) should be matched, because the carrier accumulates at the interface between the TTL and the delayed emission layer if there is an energy barrier here, which would cause a serious TPQ loss in the device. The recombination zone should be close to the delayed emission layer because the triplet excitons can be efficiently transferred from the TTL to the delayed emission layer. Additionally, the properties of the TTL are vital in achieving efficient hyper-TTF-OLEDs. As we have stated in this paper, the ideal TTL layer would have a suitable LUMO and HOMO that match the band diagram of other transport layers and the delayed emission layer, as this would reduce the primary loss and the electron- and hole-induced TPQ in TTF-OLEDs. The diffusion coefficient of the triplet energy of the TTL is also crucial. The TTL must be constructed from materials with a diffusion coefficient of at least $\rm 5\times10^{-8}$ $\rm cm^2/s$ to achieve a higher Dexter energy transfer and delayed singlet emissions. In summary, the IQE of optimized hyper-TTF-OLEDs can be improved from 29\% to 40\%, which is effectively a 33\% improvement. There is the potential to achieve more efficient blue-OLEDs by implementing the structure of hyper-TTF-OLEDs.

\medskip
\textbf{Supporting Information} \par 
Supporting Information is available from the Wiley Online Library or from the author.

\medskip
\textbf{Acknowledgements} \par 
The authors gratefully acknowledge support from the National Science and Technology Council (NSTC), Taiwan(Grant Nos.~109-2221-E-002-196-MY2, 110-2622-E-155-010 and 111-2923-E-002-009).

\medskip

%
\bibliography{TTF_AEM_manuscript}

\end{document}